\documentclass[runningheads]{llncs}
\usepackage[T1]{fontenc}
\usepackage{cite}
\usepackage{amsmath,amssymb,amsfonts}
\usepackage{graphicx}
\usepackage{textcomp}
\usepackage{xcolor}
\usepackage{booktabs}
\usepackage{lipsum}

\usepackage[numbers]{natbib}
\usepackage[ruled,vlined]{algorithm2e}
\usepackage{algpseudocode}
\usepackage{bbm}
\usepackage{mathtools}
\usepackage{caption}
\usepackage{subcaption}
\usepackage{diagbox}
\usepackage{array}
\usepackage{amsmath}
\usepackage{amssymb}
\usepackage{amsfonts}
\usepackage{mathtools}

\usepackage{breqn}
\usepackage{enumitem}
\usepackage{physics}
\usepackage{thm-restate}
\usepackage{url}
\usepackage{tikz-cd}
\newcommand{\comment}[1]{}
\allowdisplaybreaks

\def\BibTeX{{\rm B\kern-.05em{\sc i\kern-.025em b}\kern-.08em
    T\kern-.1667em\lower.7ex\hbox{E}\kern-.125emX}}
\begin{document}
%
\title{Tight Differential Privacy Guarantees for the Shuffle Model with $k$-Randomized Response}
\titlerunning{Tight Differential Privacy Guarantees for the Shuffle Model}
%
\author{Sayan Biswas \inst{1,2,3} \and
Kangsoo Jung\inst{1}\and
Catuscia Palamidessi\inst{1,2}}
\authorrunning{S. Biswas et al.}
%
\institute{Inria, Palaiseau, France \and
\'{E}cole Polytechnique, Palaiseau, France \and
EPFL, Switzerland
\\}
\maketitle              

\begin{abstract}
Most differentially private algorithms assume a central model in which a reliable third party inserts noise to queries made on datasets, or a local model where the data owners directly perturb their data. However, the central model is vulnerable via a single point of failure, and the local model has the disadvantage that the utility of the data deteriorates significantly. The recently proposed shuffle model is an intermediate framework between the central and local paradigms. In the shuffle model, data owners send their locally privatized data to a server where messages are shuffled randomly, making it impossible to trace the link between a privatized message and the corresponding sender. In this paper, we theoretically derive the tightest known differential privacy guarantee for the shuffle models with $k$-Randomized Response ($k$-RR) local randomizers, under histogram queries, and we denoise the histogram produced by the shuffle model using the matrix inversion method to evaluate the utility of the privacy mechanism. We perform experiments on both synthetic and real data to compare the privacy-utility trade-off of the shuffle model with that of the central one privatized by adding the state-of-the-art Gaussian noise to each bin. We see that the difference in statistical utilities between the central and the shuffle models shows that they are almost comparable under the same level of differential privacy protection. 
\end{abstract}


\section{Introduction}
As machine learning and data analysis using sensitive personal data are becoming more and more popular, concerns about privacy violations are also increasing manifold. The most successful approach to address this issue is differential privacy (DP). Most research performed in this area probes two main directions. One is the so-called central model, in which a trusted third party (the curator) collects the user's personal data and obfuscates them with a differentially private mechanism. The other is the local model, where the data owners apply the mechanism themselves on their data and send the perturbed data to the collector. A major drawback of the central model is that there is the risk that the curator may be corrupted. On the other hand, in the local model, there is no need to rely on a trusted curator. However, since each record is obfuscated individually, the utility of the data is substantially deteriorated compared to the central model.

In order to address the problem of the loss of utility in the local model, an intermediate paradigm between the central and the local models, known as the \emph{shuffle model (SM)} of differential privacy, was recently proposed~\cite{bittau2017prochlo}. As an initial step, the shuffle model uses a local mechanism to perturb the data individually like the local model. The difference is that, after this first step of sanitization, a shuffler uniformly permutes the noisy data to dissolve their link with the corresponding data providers. Since a potential attacker is oblivious to the shuffling process, the data providers obtain two layers of privacy protection: injection of random noise by the local randomizer and anonymity by data shuffling. This allows the shuffle model to achieve a certain level of privacy protection using less noise than the local model.

The privacy guarantees provided by the shuffle model have been rigorously analyzed in several studies. More specifically, given a local mechanism with a level of privacy parameterised by $\epsilon_0$ (pure local DP) or $(\epsilon_0,\delta_0)$ (approximate local DP), the aim is to derive a $(\epsilon,\delta)$ bound on the level of differential privacy guaranteed by applying shuffling on top of the local mechanism. In this paper, we derive the tight $(\epsilon,\,\delta)$-DP guarantee for the shuffle model with the $k$-RR local mechanism by using the concept of $(\epsilon,\,\delta)$-adaptive differential privacy (ADP) proposed by Sommer et al. in \cite{sommer2019privacy}. Next, we consider the question of how convenient the shuffle model is for publishing histograms in terms of the privacy-utility trade-off as opposed to the central model.

We perform various experiments on both synthetic and real data (the Gowalla dataset) and compare the utilities of the two models calibrated with the same privacy parameters. As expected, the utility of the central model is better than that of the shuffle model, consistent with what was observed in the literature \cite{cheu2021differential}. However, in our case, the gap is very small -- namely the histograms resulting from the shuffle model, once de-noised, are almost as close to the original ones as those of the central model. The contributions of this paper are as follows. 
\begin{enumerate}

\item we derive an analytical form of the tight differential privacy guarantee for the shuffle model with $k$-RR local randomizer under histogram queries, and therefore, show that the shuffle model, essentially, provided a higher level of DP guarantee than what is known by the community, for the same level of locally injected noise to the data.

\item using the tight bound of the $(\epsilon,\,\delta)$-DP provided by the shuffle model, as derived, we compare the privacy-utility trade-off of the shuffle model and the optimized Gaussian mechanism for the histogram queries and show that their performances are comparable. 
\end{enumerate}


\section{Related work}
Recently, intensive research on shuffle models of differential privacy has been done in various directions. One of the major research directions in this area is the study of privacy amplification by shuffling \cite{erlingsson2019amplification, balle2019privacy}.  Erlingsson et al. \cite{erlingsson2019amplification} analysed the privacy amplification of the local randomizer's privacy protection by shuffling. Balle et al.\cite{balle2019privacy} introduced the idea of privacy guarantee in shuffle models and quantitatively analyzed the relationship between the privacy parameter $\epsilon$ and the number of participants in the shuffle protocol. Feldman et al. \cite{feldman2020hiding} improved Balle et al.'s results and suggested an asymptotically optimal dependence of the privacy amplification on the privacy parameter of the local randomizer. However, neither \cite{balle2019privacy} nor \cite{feldman2020hiding} explicitly theorize any guarantee for the tightness of the bounds for the privacy guarantee of shuffle models. Koskela et al.\cite{koskela2021tight} proposed computational methods to estimate tight bounds based on weak adversaries -- however, they are not expressed by an analytical formula, they can only be computed via an algorithm. Sommer et al. introduced the notion of adapted differential privacy (ADP) in \cite{sommer2019privacy} and laid down specific conditions to achieve the tight $(\epsilon,\,\delta)$-ADP for any abstract and high-level probabilistic mechanism. To derive the tight DP guarantees for SMs, we adapt Sommer et al.'s result and obtain necessary and sufficient conditions for achieving $\delta$ that warrants the best ($\epsilon$, $\delta$)- DP guarantee in SMs with a $k$-RR local randomizer.

\section{Preliminaries}

\begin{definition}[Differential privacy\cite{DworkDifferentialPrivacy}]
\label{def:dp}
For a certain query, a randomizing mechanism $\mathcal{K}$ is \emph{($\epsilon,\,\delta)$-differentially private (DP)} if for all adjacent 
datasets, $D_1$ and $D_2$, and all $S\subseteq$ Range($\mathcal{K}$), we have:
\begin{equation*}
\mathbb{P}[\mathcal{K}(D_1) \in S] \leq e^{\epsilon}\,\mathbb{P}[\mathcal{K}(D_2) \in S] +\delta
\end{equation*}
\end{definition}

\begin{definition}[Adaptive differential privacy \cite{sommer2019privacy}]
\label{def:adp}
For $x_0,\,x_1\,\in \mathcal{X}$, where $\mathcal{X}$ is the space of the original data, and for a member $u$ in the dataset, a randomizing mechanism $\mathcal{K}$ is \emph{$(\epsilon,\,\delta)$-adaptive differentially private (ADP) for $x_0$ and $x_1$} if for all datasets, $D(x_0)$ and $D(x_1)$, and all $ S \subseteq$ Range($\mathcal{K}$), we have:
\begin{equation*}
\mathbb{P}[\mathcal{K}(D(x_0)) \in S] \leq e^{\epsilon}\,\mathbb{P}[\mathcal{K}(D(x_1) \in S]\,+\,\delta
\end{equation*}
where $D(x_0)$ and $D(x_1)$ are datasets differing only in the entry of the fixed member $u$: $D(x)$ means that $u$ reports $x$ for every $x\,\in\,\mathcal{X}$, keeping the entries of all the other users the same.
\end{definition}

\begin{remark}\label{ADPandDP}
$\mathcal{K}$ is $(\epsilon,\,\delta)$-DP implies that $\mathcal{K}$ is $(\epsilon,\delta)$-ADP for every $x_0,x_1\in\mathcal{X}$.   
\end{remark}

\begin{definition}[Tight DP (or ADP) \cite{sommer2019privacy}]\label{def:tight}
    Let $\mathcal{K}$ be $(\epsilon,\,\delta)$-DP (or ADP for $x_0,\,x_1\in\mathcal{X}$). We say that $\delta$ is \emph{tight} for $\mathcal{K}$ (w.r.t. $\epsilon$ and $x_0,\,x_1$ in case of ADP) if there is no $\delta'<\delta$ such that $\mathcal{K}$ is $(\epsilon,\,\delta')$-DP (or ADP for $x_0,\,x_1$).
\end{definition}

\begin{definition}[Local differential privacy\cite{duchi2013local}]
\label{def:ldp}
Let $\mathcal{X}$ denote a possible alphabet for the original data and let $\mathcal{Y}$ be the alphabet of noisy data.  A randomizing mechanism $\mathcal{R}$ provides \emph{$\epsilon$-local differential privacy (LDP)} if for all $x_1,\,x_2\,\in\,\mathcal{X}$, and all $y\,\in\,\mathcal{Y}$, we have
\begin{equation*}
\mathbb{P}[\mathcal{R}(x_1)=y] \leq e^{\epsilon}\,\mathbb{P}\left(\mathcal{R}(x_2)=y\right)
\end{equation*}
\end{definition}

\begin{definition}[k-Randomized Response\cite{kairouz2016discrete}]
\label{def:krr}
Let $\mathcal{X}$ be a discrete alphabet of size $k$. Then \emph{k-randomized response} ($k$-RR) mechanism, $\mathcal{R}_{\text{kRR}}$, is a locally differentially private mechanism that stochastically maps $\mathcal{X}$ onto itself (i.e., $\mathcal{Y}=\mathcal{X}$), given by
\begin{equation*}
    \mathcal{R}_{\text{kRR}}(y|x)=
        \begin{cases}
		    c\,e^{\epsilon} & \text{, if $x=y$}\\
            c, & \text{, otherwise}
		\end{cases}
\end{equation*}
for any $x,\,y\,\in\mathcal{X}$, where $c=\frac{1}{e^{\epsilon}+k-1}$.

\end{definition}

\begin{definition}[Shuffle model\cite{erlingsson2019amplification}]
\label{def:shuffleseq}
Let $\mathcal{X}$ and $\mathcal{Y}$ be discrete alphabets for the original and the noisy data respectively. For any dataset of size $n\,\in\,\mathbb{N}$, the \emph{shuffle model (SM)}  is defined as $\mathcal{M}:\mathcal{X}^n\mapsto\mathcal{Y}^n$, $\mathcal{M}=\mathcal{S} \circ \mathcal{R}^n$, where
\begin{itemize}
    \item $\mathcal{R}:\mathcal{X}\mapsto\mathcal{Y}$ is a local randomizer, stochastically mapping each element of the input dataset, sampled from $\mathcal{X}$, onto an element in $\mathcal{X}$, providing $\epsilon_0$-local differential privacy.
    \item $\mathcal{S}:\mathcal{Y}^n\mapsto\mathcal{Y}^n$ is a shuffler that uniformly permutes the finite set of messages of size $n\,\in\,\mathbb{N}$, that it takes as an input.
\end{itemize} 
\end{definition}

A SM can be perceived as having a sequence of messages going through the mechanism $\mathcal{M}$ and then coming out as the frequencies of each of the noisy messages, as the idea of the layer of ``shuffling'' is to randomize the noisy messages w.r.t. their corresponding senders by a random permutation. Let us call this particular brand of query on SM as the \emph{histogram query}.

\begin{definition}[Histogram query \cite{balcer2019separating}]\label{def:shufflehist}
Let $\mathcal{X}$ and $\mathcal{Y}$ be discrete alphabets for the original and the noisy data respectively. For any dataset of size $n\,\in\,\mathbb{N}$, the \emph{histogram query} on SM, $\mathcal{M}:\mathcal{X}^n\mapsto\mathbb{R}^{+\,n}$, is defined as $\mathcal{M}=\mathcal{T} \circ \mathcal{R}^n$, where
\begin{itemize}
    \item $\mathcal{R}:\mathcal{X}\mapsto\mathcal{Y}$ is a local randomizer providing $\epsilon_0$-local differential privacy, as in Definition \ref{def:shuffleseq}.
    \item $\mathcal{T}:\mathcal{Y}^n\mapsto\mathbb{R}^n$ is a function that gives the frequency of each message in finite set of messages of size $n\,\in\,\mathbb{N}$, that it takes as an input.
\end{itemize} 
In other words, if we have a dataset $D_{\mathcal{X}}=(x_1,\ldots,x_n)\in\mathcal{X}^n$, then $D_{\mathcal{Y}}=\mathcal{M}(D_{\mathcal{X}})=\mathcal{T}((\mathcal{R}(x_1),\ldots,\mathcal{R}(x_n))=(s_{1},\ldots,s_{n})$, where $s_i=n_i/n$ with $n_i$ denoting the number of times $\mathcal{R}(x_i)$ occurs in $D_{\mathcal{Y}}$.
\end{definition}

\begin{definition}[Privacy loss random variable \cite{sommer2019privacy}]\label{def:PLrv}
For a probabilistic mechanism mapping messages from the alphabet of original messages to the alphabet for noisy messages, $M:\mathcal{X}\mapsto\mathcal{Y}$, let us fix $x_0,\,x_1\,\in\,\mathcal{X}$ and a potential output $y\,\in\,\mathcal{Y}$. The \emph{privacy loss random variable} of $y$ for $x_0$ over $x_1$ is defined as:
where $M(x_i)$ is the probability distribution of the noisy output for the original input $x_i$ for $i\in\{0,1\}$.

\begin{align}
 \mathcal{L}_{M(x_0)/M(x_1)}(y)
    =\begin{cases}
		    +\infty & \begin{cases}
		                 & \mathbb{P}(M(x_0)=y)\neq0,\\
		                & \mathbb{P}(M(x_1)=y)=0
		    \end{cases}\\
            \ln{\frac{\mathbb{P}(M(x_0)=y)}{\mathbb{P}(M(x_1)=y)}} &
            \begin{cases}
		                 & \mathbb{P}(M(x_0)=y)\neq0,\\
		                & \mathbb{P}(M(x_1)=y)\neq0
		    \end{cases}\\
            -\infty & \text{o.w.}
		\end{cases}
\end{align}
\par

\end{definition}

\begin{definition}[Privacy loss distribution \cite{sommer2019privacy}]\label{def:PLD}
Let $P_1$ and $P_2$ be two probability distributions on $\mathcal{Y}$ (the finite alphabet for noisy messages). The \emph{privacy loss distribution, $\omega$, for A over B} is defined as:
\begin{equation*}
    \omega(u)=\sum\limits_{y:\mathcal{L}_{A/B}(y)=u}\mathbb{P}(A=y) \text{ for all $u\in\mathcal{U}$, where $\mathcal{U}=\bigcup\limits_{y\in\mathcal{Y}}\{\mathcal{L}_{A/B}(y)\}\subset \mathbb{R}$.}
\end{equation*}

\end{definition}

 

\section{Tight privacy guarantee for SM}

\subsection{Overview}\label{outline}
Sommer et al. in \cite{sommer2019privacy} proposed a notion of adaptive differential privacy and derived a very important sufficient and necessary result for any probabilistic mechanism to have the best formal privacy guarantee. Adaptive differential privacy essentially translates the idea of a differential privacy guarantee with respect to a chosen pair of elements in the dataset. Exploiting this result (Result 1), we derived the necessary and sufficient condition needed to warrant the best DP guarantee for SM with the most popularized LDP satisfying local randomizer, the $k$-RR mechanism. This essentially draws the tight DP guarantee that an SM can induce being locally randomized with a $k$-RR mechanism. At the crux of this paper, the importance of deriving the tight DP guarantee by SM under the $k$-RR local randomizer implies that we show that the SM provides a higher level of privacy than what is known by the existing work in the literature that focuses on improving the privacy bound for the SM.  
\setlength\intextsep{\glueexpr\intextsep/2\relax}
\begin{table} 
\caption{Value of $\delta$ derived from the existing work and our proposed result}\label{table:tab1}
 \centering
  \begin{tabular}{cccccc}
    \hline
    & \cite{erlingsson2019amplification} &  \cite{balle2019privacy} & \cite{feldman2020hiding}  & 
    \cite{koskela2021tight} &
    Proposed tight $\delta$ \\
    \hline
    $\epsilon=0.1$ & 0.97 &	0.229 & 0.066 & 9.01E-4 &2.38E-28 \\
    \hline
    $\epsilon=0.2$ & 0.89 &	0.002 &	1.91E-5 & 1.89E-6  &1.61E-42 \\
    \hline
    $\epsilon=0.3$ & 0.77&	1.77E-6 & 2.43E-11 &  2.19E-10 & 5.22E-57 \\
    \hline
    $\epsilon=0.4$ & 0.64 &	5.95E-11 & 1.35E-19 &3.14E-16 &	5.14E-72 \\
    \hline
  \end{tabular}%
 
\end{table}
\setlength\intextsep{\glueexpr\intextsep/2\relax}

Table \ref{table:tab1} presents the values of $\delta$ obtained from the results in \cite{erlingsson2019amplification}, \cite{balle2019privacy}, \cite{feldman2020hiding},\cite{koskela2021tight} and the proposed derivation in \eqref{eq:tightdelta} of this paper, by varying $\epsilon$ from $0.1$ to $0.4$, fixing $n=100$ and $\epsilon_0=0.5$. We observe that, indeed, the value of $\delta$ computed from \eqref{eq:tightdelta} in Definition \ref{finaldelta} is significantly less compared to the other existing improvements proposed, highlighting that our proposed result engenders the best possible DP guarantee for SMs under the $k$-RR local randomizer.

\subsection{Framework}\label{framework}
Let $\mathcal{X}=(x_0,\ldots,x_{k-1})$ be the alphabet of messages of size $k\in\mathbb{N},\,k>1$ and $\mathfrak{U}$ be the set of all users involved in the environment. For simplicity, we assume the alphabets of the original and noisy messages to be the same, both being $\mathcal{X}$. Therefore, the local randomizer of our shuffle mechanisms locally sanitizes the dataset by mapping original messages sampled from $\mathcal{X}$ to elements of $\mathcal{X}$. 

Let $\epsilon_0$ be the privacy parameter of $\mathcal{R}_{\text{kRR}}$, which is used as the local randomizer for the shuffle mechanisms discussed in this paper. Furthermore, letting $D_{\mathcal{X}}$ be the dataset of the original messages of $n$ users, each of which is sampled from (and obfuscated to) $\mathcal{X}$, we denote $D_{\mathcal{X}\,z}$ as the original message of $z\in\mathfrak{U}$ in $D_{\mathcal{X}}$ for any $z\in\mathfrak{U}$.  Let $D_{\mathcal{Y}}=\mathcal{R}_{\text{kRR}}^n(D_{\mathcal{X}})=\{\mathcal{R}_{\text{kRR}}(D_{\mathcal{X}\,z}):z\in\mathfrak{U}\}$ be the noisy dataset going through $\mathcal{R}_{\text{kRR}}$.

For the purpose of analysing the adaptive differential privacy, let us fix a certain user, $u\in\mathfrak{U}$, whose data is in $D_{\mathcal{X}}$. Since the only major distinction that $k$-RR mechanism makes in the process of mapping a datum from its original value to the obfuscated value is whether the original value and the obfuscated value are the same or not (i.e., the probability that the $x$ is being reported as $x'$ is the same for every $x'\in\mathcal{X}$ when $x\neq x'$), it is reasonable for us to study the adaptive differential privacy guarantee with respect to a couple of potential original messages of $u$, say $x_0,\,x_1\in\mathcal{X}$, $x_0\neq x_1$ in the environment where the shuffle model uses a $k$-RR local randomizer. 

The idea behind adaptive differential privacy w.r.t. $x_0,\,x_1$ is to make it significantly difficult to predict whether $u$'s original message is $x_0$ or $x_1$. In the context of this work, since we will be focusing on the case of having the local randomizer as the $k$-RR mechanism, the only gravity $x_1$ holds as far as the shuffle model is concerned is the fact that it is different from $x_0$. Thus $x_1$ could represent any $x\in\mathcal{X}$ such that $x\neq x_0$. Therefore, we shall be analysing the privacy of $u$'s original message being $x_0$ and compare its privacy level of being identifiable with a different potential original message, which we fix as $x_1$ w.l.o.g. Let's call $x_0$ as the \emph{primary input} for $u$ and $x_1$ be the \emph{secondary input}. For a fixed set of values reported by every user in $\mathfrak{U}\setminus\{u\}$, let $D(x_0)$ represent the edition of the dataset where $u$ reports $x_0$, and let $D(x_1)$ represent the one where $u$ reports $x_1$. 

The most important result from literature - Lemma 5 in \cite{sommer2019privacy} - that is heavily exploited in this paper is as follows:

\textbf{\emph{Result 1:}} (Lemma 5 \cite{sommer2019privacy})
For every probabilistic mechanism $M:\mathcal{X}\mapsto\mathcal{Y}$, for any $x_0,\,x_1\,\in\,\mathcal{X}$ and any $\epsilon,\,\delta(\epsilon)>0$, $M$ is tightly $(\epsilon,\,\delta)$-ADP for $x_0,\,x_1$ iff 
\begin{align}\label{tightdeltacondition}
    \textstyle\delta(\epsilon)
    = \omega(\infty)+\sum\limits_{\substack{u\in \mathcal{U}\setminus \{\infty, -\infty\}  \\ u>\epsilon}} (1-e^{\epsilon-u})\omega(u)
\end{align}

\subsection{Theorems and results}
 
As we are interested in finding $\epsilon>0$ and, correspondingly, $\delta>0$ that provide a tight ADP guarantee for $\mathcal{M}$ for $x_0,\,x_1$, we define the constants $\kappa_1,\kappa_2,\kappa_3$ to simplify the mathematical results derived in the subsequent sections as follows:

\begin{align}
    &\kappa_1\coloneqq\frac{e^{\epsilon_0}(e^{\epsilon_0}+k-2)}{k-1}\label{kappa6}\\
    &\kappa_2\coloneqq\frac{k-1}{e^{\epsilon_0}+k-2}\label{kappa7}\\
    &\kappa_3\coloneqq\frac{(k-1)^{n_{x_0}}(e^{\epsilon_0}+k-2)^{n-n_{x_0}-s}}{(e^{\epsilon_0}+k-1)^n}\label{kappa8}
\end{align}




\begin{remark}\label{rem:positiveconstants}
Note that $\kappa_1,\kappa_2,\kappa_3>0$ for any $\epsilon_0>0$, $n\in\mathbb{N},\,k\in\mathbb{N}_{\geq 2},\, s\in\mathbb{N}$.
\end{remark}

From now on we shall focus on the histogram query of the shuffle model. For the same $\epsilon_0$-LDP mechanism $\mathcal{R}_{\text{kRR}}$ to be used as the local randomizer for histogram query, let $\mathcal{M}=\mathcal{T}\circ\mathcal{R}_{\text{kRR}}$ denote the shuffle model that takes in a sequence of original messages, obfuscates them locally using $\mathcal{R}_{\text{kRR}}$, and broadcasts the frequency of each message in the noisy dataset. In other words, having $u$ having $x_i$ as her original message for $i\in\{0,\,1\}$, $\mathcal{M}(D(x_i))=(\mathcal{M}_{x_0}(x_i),\ldots,\mathcal{M}_{x_{k-1}}(x_i))$ where $\mathcal{M}_{x_j}(x_i)$ is a random variable giving the frequency of $x_j\in \mathcal{X}$ in the noisy dataset, $D_{\mathcal{Y}}$, obfuscated by $\mathcal{R}_{\text{kRR}}$. Assuming that $u$'s original data is $x_0$ (w.l.o.g.), let $n_{x_0}$ denote the number of times $x_0$ has appeared in $D_{\mathcal{X}}$ for the original entries from all users in $\mathfrak{U}\setminus u$.

\begin{definition}\label{PLrvHistQuery}
    By Definition \ref{def:PLrv}, the \emph{privacy loss random variable for the histogram query for shuffle model} of $x_0$ over $x_1$ with respect to a certain output $s\in\mathbb{N}$, in $\mathcal{M}$ is $v_s(x_0,x_1)=\ln{\frac{\mathbb{P}(\mathcal{M}_{x_0}(x_0)=s)}{\mathbb{P}(\mathcal{M}_{x_0}(x_1)=s)}}$.
\end{definition}
\begin{definition}\label{finaldelta}
    For $s \in \{0,\ldots,n\}$, $r\in\{0,\ldots,s\}$, let $\mu(s,r)=\binom{n_{x_0}}{r}\binom{n-n_{x_0}}{s-r}\kappa_1^r$ and $\tau_r=\kappa_2(n-n_{x_0})+(e^{\epsilon_0}-\kappa_2)(s-r)$.
    For any $\epsilon>0$, let us define
    \begin{equation}\label{eq:tightdelta}
        \hat{\delta}(\epsilon)\coloneqq\sum\limits_{s=0}^n\mathbbm{1}_{\{v_{s}(x_0,x_1)>\epsilon\}}(1-e^{\epsilon-v_{s}(x_0,x_1)})\frac{\kappa_3}{n-n_{x_0}}\sum\limits_{r=0}^s\mu(s,r)\tau_r
    \end{equation}
where $\mathbbm{1}_{E}$ is the indicator function for any event $E$.
\end{definition}


\begin{restatable}{theorem}{shufflemodelprobfreqquery}\label{th:tightdelta_ADP}
For any $\epsilon>0$, we get the tight $(\epsilon,\,\delta)$-ADP guarantee for $\mathcal{M}$ with respect to $x_0,\,x_1$ iff $\delta=\hat{\delta}(\epsilon)$ as in as in \eqref{eq:tightdelta} of Definition~\ref{finaldelta} where $$v_s(x_0,x_1)=\ln{\left(\kappa_2+\frac{\frac{(e^{\epsilon_{0}}-\kappa_2)}{n-n_{x_0}}\left(\sum\limits_{r=0}^{s}(s-r)\binom{n_{x_0}}{r}\binom{n-1-n_{x_0}}{s-1-r}\kappa_1^r\right)}{\sum\limits_{r=0}^{s}\binom{n_{x_0}}{r}\binom{n-n_{x_0}}{s-r}\kappa_1^r}\right)}.$$
\end{restatable}

\begin{restatable}{corollary}{shufflemodelDP}\label{th:tightdelta_DP}
For any $\epsilon>0$, we get the tight $(\epsilon,\,\delta)$-DP guarantee for $\mathcal{M}$ iff:
\begin{equation}
    \delta(\epsilon)\coloneqq\sum\limits_{s=0}^n\mathbbm{1}_{\{v_s>\epsilon\}}(1-e^{\epsilon-v_s})\frac{\kappa_3}{n-
    n_{x_0}}\sum\limits_{r=0}^s\mu(s,r)\tau_r
\end{equation}
where
$v_s=\max_{x_0,x_1\in\mathcal{X}}v_s(x_0,x_1)$ and $v_s(x_0,x_1)$ is as derived in Theorem~\ref{th:tightdelta_ADP}.
\end{restatable}

\section{Evaluating the utility of the shuffle model}

It is crucial to have the tight bound in the privacy guarantee for shuffle models to be able to conduct a fair comparison of utilities of shuffle models with other forms of differential privacy under a certain level of privacy protection. 

Suppose with $\epsilon, \,\delta$, we get a tight $(\epsilon,\,\delta)$-ADP guarantee for $\mathcal{M}$ w.r.t. $x_0$ as the primary input. We wish to compare how the utility of $\mathcal{M}$ would perform against that of a central model of differential privacy for histogram query implemented on $D_{\mathcal{X}}$ with the same privacy parameters $\epsilon$ and $\delta$. For this, we will be sticking to the most optimal framework, known until now \cite{balle2018improving}, of one of the most popular mechanisms for the central model for $(\epsilon,\,\delta)$-DP: the \emph{Gaussian mechanism}. The details of the theoretical build-up are provided in Appendix~\ref{app:theory_outline}.

In \cite{cheu2019distributed}, Cheu et al. give theoretical evidence that the accuracy of the SM lies in between the central and local models of DP. However, no experimental analysis had been performed to dissect how low the accuracy of SMs lies when compared to the central model when both provide the same level of privacy protection. Thus, the main goal of our experiments was to empirically show the scale of difference in accuracy between SM and the central model by comparing their statistical utilities under the tight and equal DP guarantee. To do this end, we compared the statistical approximation of the true distribution from the SM with $k$-RR local randomizer to that of the central model by applying the Gaussian mechanism \cite{balle2018improving}, using the value of $\delta$ derived from \eqref{eq:tightdelta}, ensuring the tight $(\epsilon, \delta)$-DP guarantee.

\subsection{Experimental results on synthetic data}
In this section, we carry out an experimental analysis to illustrate the comparison of utilities for histogram query of the shuffle model using $k$-RR local randomizer and the optimal Gaussian mechanism using synthetically generated data sampled from $\mathcal{N}(0,2)$. We experimented and demonstrated our results in the two categories: (i) trend analysis of $\delta$ providing the tight ADP guarantee for $\mathcal{M}$ and (ii) utility comparison between $\mathfrak{N}$ and $\mathcal{M}$ under the same level of differential privacy. 

To analyze the values of $\delta$ providing a tight ADP guarantee for $\mathcal{M}$, we change the values of $\epsilon$, $\epsilon_0$, $n$, $n_0$, and $k$ that enable us to see the change in the trend of $\delta$. For comparing the utilities of the central model and the shuffle model, we considered $\hat{\delta}$ as in \eqref{eq:maxdelta}, providing the worst possible tight ADP over every $x_0,x_1\in\mathcal{X}$, and therefore, by Remark \ref{ADPandDP}, a DP guarantee. Table \ref{table:tab0} shows the default values of the parameters used for the experiment.

\begin{table}
\caption{Experimental parameters used for synthetic data}\label{table:tab0}
 \centering
\begin{tabular}{cc}
\hline
 Parameter name & Values \\ \hline
$\epsilon$  & 0.1 to 3 \\
$\epsilon_0$ & 0.1 to 3 \\
$n$ & 50, 100, 150, 1000, 100000 \\
$x_0$ & 1 to 15\\
$k$ & 5, 10, 15
 \\ \hline
\end{tabular}
\end{table}

\subsubsection{Tight $\delta$ for histogram queries}

We show the experimental results for deriving $\delta$ providing ($\epsilon,\,\delta$)-ADP guarantee, as given by Theorem \ref{th:tightdelta_ADP}, by changing the values for $\epsilon$, $\epsilon_0$, $n$ and $k$. We use the \emph{total variation distance}, $d_{TV}(.)$, to evaluate $\mathcal{W}(\mathcal{M})$ and $\mathcal{W}(\mathfrak{N})$ -- the ``distances'' of the estimated original distribution obtained from shuffle model with $k$-RR local randomizer, using matrix inversion, (shuffle+INV), and the distribution sanitized with Gaussian mechanism from the original distribution itself. Table \ref{table:tight_delta} shows $\delta$ when we vary $\epsilon$,  for three categories:
\begin{enumerate}
    \item[(a)] We change $\epsilon_0$, fixing $n_{x_0}=80$, $n=100$, and $k=10$. We observe that $\delta$ decreases as $\epsilon$ increases for the same $\epsilon_0$, and $\delta$ increases as $\epsilon_0$ increases under a fixed value of $\epsilon$. When it does not satisfy the $v_s>\epsilon$ condition of equation (57), $\delta$ becomes 0. For a fixed $\epsilon$ and $\epsilon_0$, a high value of $\delta$ decreases the level of privacy protection. Thus, experimentally, we can validate that for a constant $\epsilon$, $\delta$ increases as $\epsilon_0$ used for $k$-RR increases, ensuring that the privacy protection of the shuffle model decreases with a decrease in the privacy level of its local randomizer.
    \item [(b)] We vary $n$ fixing $k=10$, $\epsilon_0=2$, and $n_{x_0}=80$. For the same $\epsilon$, $\delta$ becomes smaller as the value of $n$ increases. A lower $\delta$ means higher privacy protection, reassuring that the shuffle model provides higher privacy protection as the number of users (samples) increases.
    \item[(c)] We alter $k$ fixing $n=100$, $\epsilon_0=2$, and $n_{x_0}=80$. As the value of $k$ increases, $\delta$ decreases. This is also due to the characteristic of the $k$-RR mechanism, which is used as the local randomizer for $\mathcal{M}$. The inference probability for a potential adversary decreases as the size of the domain for the data increases.
\end{enumerate}

\begin{table}
\caption{Tight $\delta$ for different $\epsilon$}\label{table:tight_delta}
\centering
\begin{tabular}{cccccccc}
\hline
&&& Varying $\epsilon_0$ &&&&\\
\hline
 & $\epsilon=0.1$& $\epsilon=0.5$ & $\epsilon=1.0$ & $\epsilon=1.5$ & $\epsilon=2.0$ & $\epsilon=2.5$ & $\epsilon=3.0$  \\ \hline
$\epsilon_0=1$ &2.08E-20
  & 3.42E-43
 & 0 & 0 & 0 & 0 & 0 \\
$\epsilon_0=2$ & 2.49E-15
  & 3.25E-22
 & 2.20E-30
  & 1.57E-40
 & 0 & 0 & 0 \\
$\epsilon_0=3$ & 8.79E-11
 & 5.73E-13
 & 3.52E-16
  & 4.49E-20
 & 1.09E-25
 & 4.52E-33
 & 0 \\
\hline
&&& Varying $n$ &&&&\\
\hline
 $n=$50 &1.91E-08
  & 5.02E-12
 & 2.40E-15
 & 4.49E-21
 & 0 & 0 & 0 \\
$n=$100 & 2.49E-15
  & 3.25E-22
 & 2.20E-30
  & 1.57E-40
 & 0 & 0 & 0 \\
$n=$150 & 6.58E-22
 & 7.83E-32
 & 2.75E-44
  & 6.99E-59
 & 0
 & 0
 & 0 \\
\hline
&&& Varying $k$ &&&&\\
\hline
 $k=$5 &1.96E-10
  & 2.51E-14
 & 1.08E-19 & 2.02E-27 & 0 & 0 & 0 \\
$k=$10 & 2.49E-15
  & 3.25E-22
 & 2.20E-30
  & 1.57E-40
 & 0 & 0 & 0 \\
$k=$15 & 1.66E-18
 & 7.13E-28
 & 1.49E-38
  & 7.35E-50
 & 0
 & 0
 & 0 \\ \hline
\end{tabular}

\end{table}

\subsubsection{Comparing the utility of the shuffle  and the central models}

In this section, we compare the utilities of the central model and the shuffle models, providing the same level of privacy protection. For neutral comparison, we perform the experiments into two cases: individual specific utility and community level utility, as described in Appendix~\ref{app:theory_outline}. We use the \emph{total variation distance} to estimate the difference between original distribution and estimated distributions.

\begin{table}[ht]\centering 

\caption{Individual specific utility comparison of central and shuffle models for synthetic data ($\epsilon=4$)}\label{table:syn_individual}

  \begin{tabular}{ccc|cc}
    \hline
    {$x_0$} &
      \multicolumn{2}{c|}{$n=1,000$} &
      \multicolumn{2}{c}{$n=100,000$}\\
    & Gaussian &  shuffle+INV & Gaussian  & shuffle+INV \\
    \hline
    1 & 3E-3 & 1E-3 & 6E-6 & 3E-3 \\
    \hline
    3 & 6E-4 & 2E-4 & 1E-5 & 5E-4 \\
    \hline
    5 & 12E-4 & 11E-4 & 1E-5 & 5E-4 \\
    \hline
    7 & 1E-4 & 4E-3 & 8E-6 & 3E-5 \\
    \hline
  \end{tabular}%
 
\end{table}

Table \ref{table:syn_individual} shows the results from the experimental analysis of comparing the individual specific utilities of $\mathcal{M}$ and $\mathfrak{N}$ as the primary input, $x_0$, is changed. We performed the experiments for the case of $n=1,000$ and $n=100,000$, setting $\epsilon_0=4$, $\epsilon=4$, and $k=15$, calculating $\delta$ for each $x_0$. When $n=1,000$, shuffle+INV is comparable with the Gaussian mechanism, depending on the value of $x_0$. However, when $n$ is $100,000$, the Gaussian mechanism shows better results regardless of $x_0$. This is explained through our choice of $\delta$ (given by Theorem \ref{th:tightdelta_DP}), which depends on $n_{x_0}$, which, in turn, varies with $x_0$ and that Gaussian mechanism inserts fixed noise regardless of $n$. However, even for a large value of $n$, the utility of shuffle+INV, although slightly worse than the Gaussian mechanism, is quite good as $\overline{\mathcal{W}}(\mathcal{M},x_0)$ remains very low across different $x_0$.

For the community level utility, we apply the worst case (highest value) of $\delta$ computed over all the primary inputs for all the users in $\mathfrak{U}$, given as $\hat{\delta}$ in \eqref{eq:maxdelta}, to sanitize all input messages of the dataset -- thus establishing the worst tight ADP guarantee possible on the shuffle model. This is used to determine the community level utility of the corresponding shuffle model with the estimated differential privacy guarantee. Similar to the case of individual specific utility, experiments were performed for the case of $n=1,000$ and $n=100,000$, and the other parameters used for the experiment being the same.  The experiments results are similar to what we showed for individual specific utility. When $n$ is small, the utility of shuffle model is almost as much as that of the central model. As $n$ increases, the utility of the the Gaussian mechanism, $\mathfrak{N}$, improves slightly over that of the shuffle model under the same level of differential privacy, however they still are fairly close.

\subsection{Experimental results on real data}

\begin{figure}[htbp]
\centerline{\includegraphics[width=0.6\textwidth]{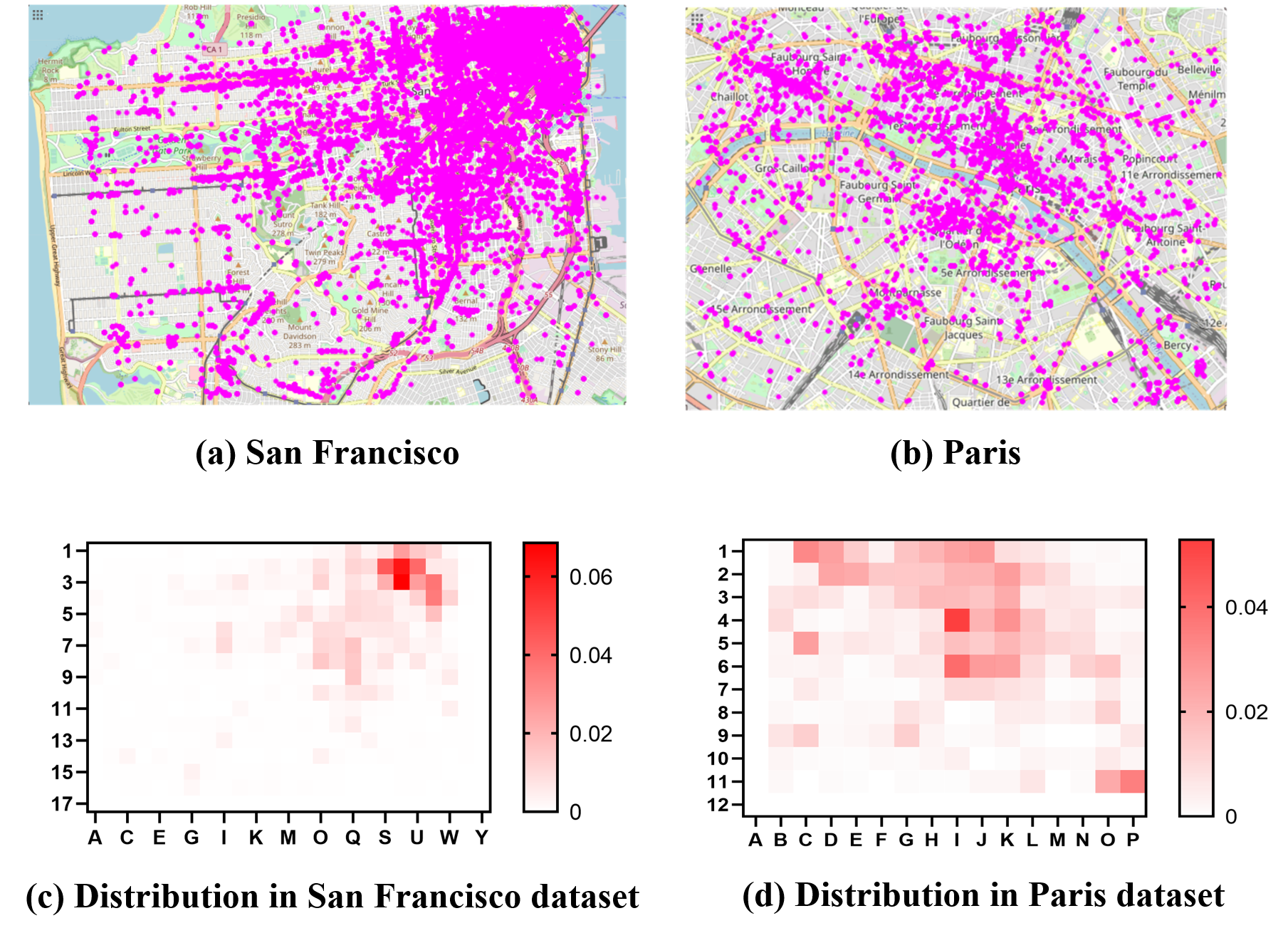}}
\caption{(a) and (b): Location data from Gowalla check-ins from a northern part of San Francisco and a part of Paris. (c) and (d) give the heatmap of the locations in the areas of San Francisco and Paris as an alternative visualization.}
\label{san&paris original}
\end{figure}

\begin{figure}[htbp]
     \centering
     \begin{subfigure}[b]{0.45\textwidth}
         \centering
         \includegraphics[width=\textwidth]{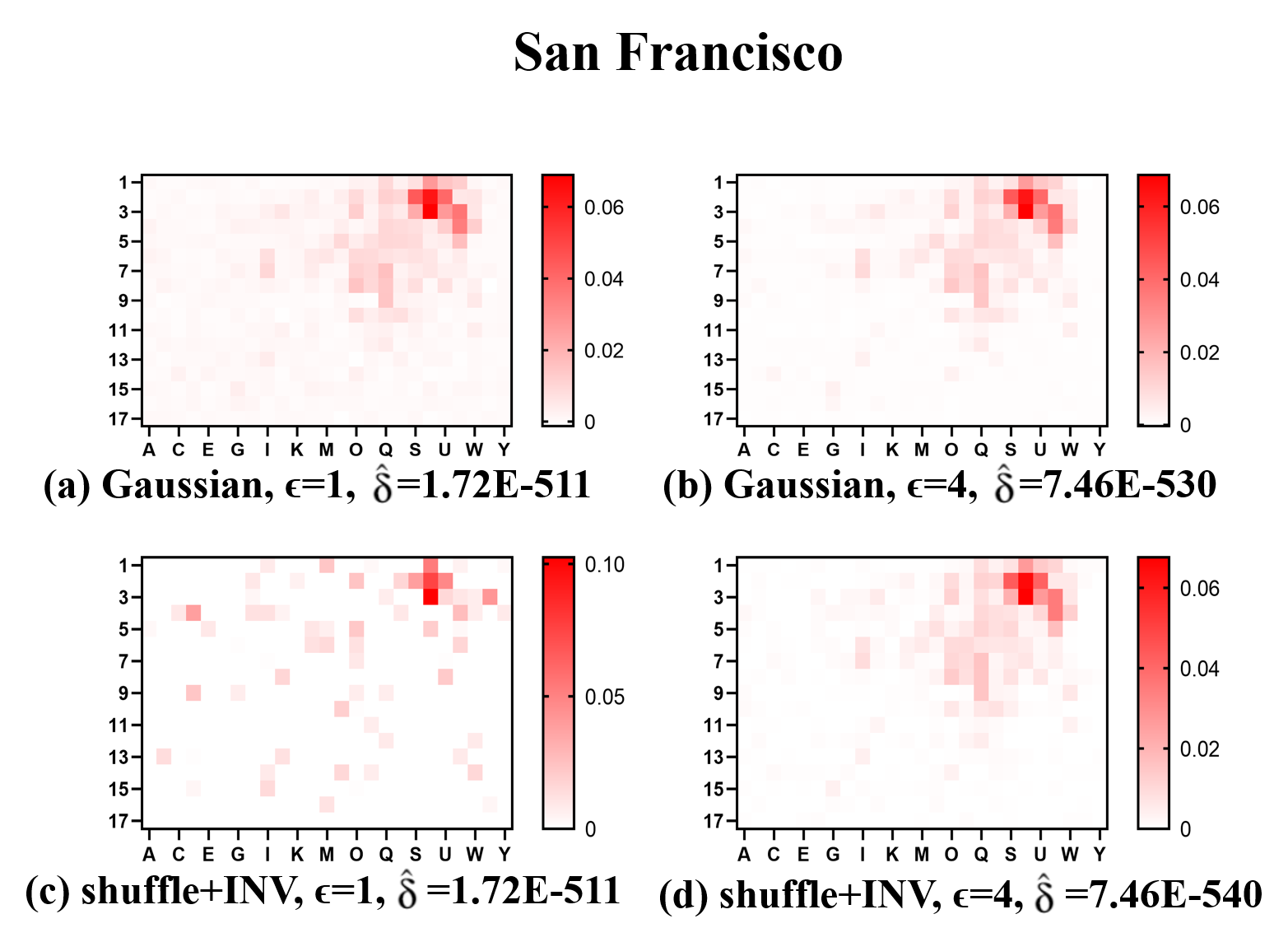}
     \end{subfigure}
     \centering
     \begin{subfigure}[b]{0.45\textwidth}
         \centering
         \includegraphics[width=\textwidth]{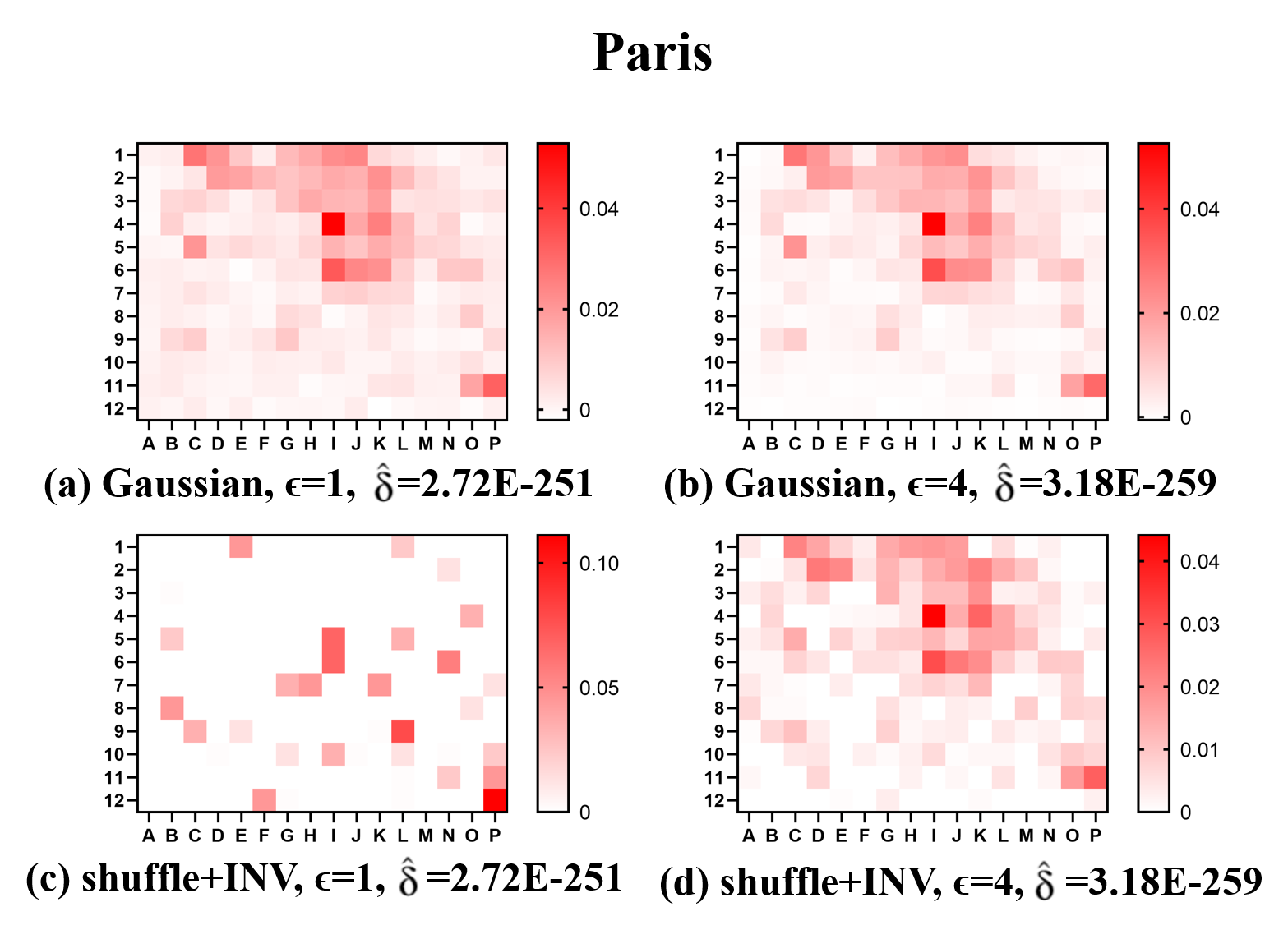}
     \end{subfigure}
     \hfill
\caption{Estimation of the original distribution from the noisy data obfuscated with the Gaussian mechanism and the SM in San Francisco and Paris dataset}
\label{san_paris_noise}
\end{figure}

Now we focus on the experimental results obtained using real location data from the Gowalla dataset\cite{Gowalla:online}. Figures \ref{san_paris_noise} illustrate the estimations of the original distributions of location data from San Francisco and Paris, respectively. We sanitize the original distribution using the shuffle model giving a tight differential privacy guarantee with parameters $\epsilon$ and $\hat{\delta}$, as in \eqref{eq:maxdelta}. We use the same $\epsilon$ and $\hat{\delta}$ to privatize the original data using the Gaussian mechanism as same in the previous experiment, thus getting a $(\epsilon,\,\hat{\delta})$-DP guarantee for both cases.

To compare the utility of the two mechanisms under the same privacy level, we estimate the original distributions using shuffle+INV for the shuffle model and the Gaussian mechanism itself for the central model, as described in \eqref{commintyutility:shuffle} and \eqref{communityutility:central} and evaluate how far the corresponding estimations lie from the original distributions.  We observe that the Gaussian mechanism approximates the original distributions slightly better than the shuffle+INV, but they are comparable. 

\begin{figure}[htbp]
\centering
\begin{subfigure}[b]{0.60\textwidth}
\centerline{\includegraphics[width=1\textwidth]{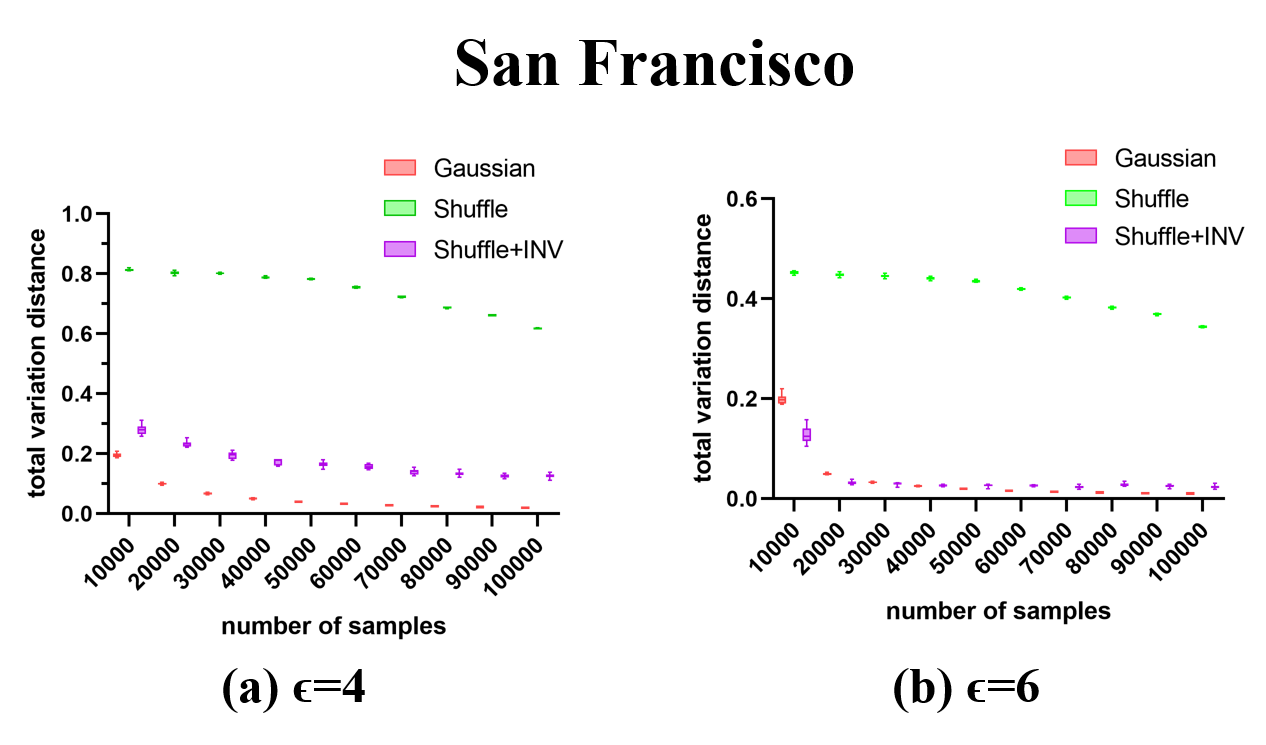}}
\end{subfigure}
\begin{subfigure}[b]{0.6\textwidth}
\centerline{\includegraphics[width=1\textwidth]{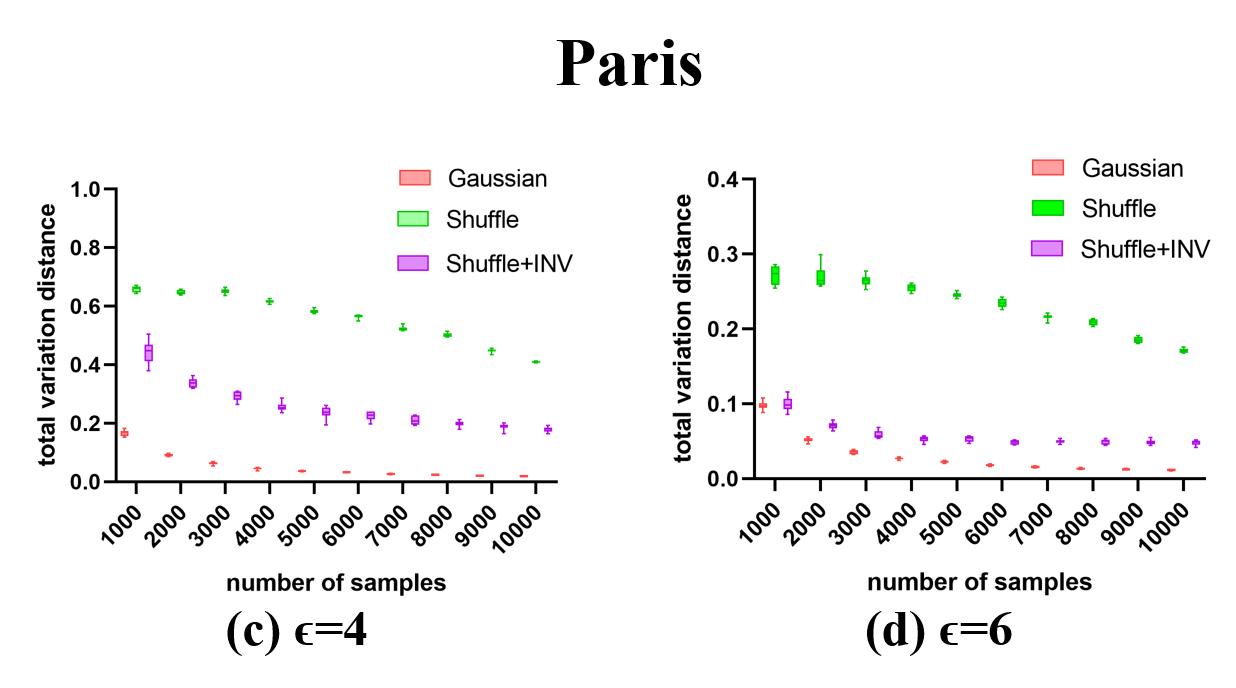}}
\end{subfigure}
\caption{Illustrating the comparison of community level utilities between Gaussian, shuffle and shuffle+INV for varying $n$ and $\epsilon$ in San Francisco and Paris dataset }
\label{san_paris_box}
\end{figure}

As we observe in the previous experiment results, the number of samples, $\epsilon$ affects the utility. In Figures \ref{san_paris_box}, we show how the number of samples and the differential privacy parameters affect the utilities in more detail. In summary, we observe a consistency with the existing work in the trend of the Gaussian mechanism having a better utility than the shuffle model across all settings. However, when the number of samples is small and the privacy level is low, the utilities of the shuffle model and the central model are comparable.

Figures \ref{san_paris_box} (a) and (b) illustrate the evaluation of the TV distance between the original and the estimated distributions for San Francisco dataset. $n$ ranges from $10,000$ to $100,000$, which is used to sample locations from the aforementioned San Francisco region. We set $\epsilon=4$ and $\epsilon=6$ to capture the change of distance between the original and the estimated distributions by varying $n$. We use $\hat{\delta}$, as in \eqref{eq:maxdelta}, to calculate community-level utility and we run the mechanism 10 times to obtain the boxplots. The results exhibit that shuffle model$, \mathcal{M}$, gives worse utility than the central model $\mathfrak{N}_{\epsilon,\hat{\delta}}$, and shuffle+INV shows better utility than shuffle. This trend is harmonious across the different settings for $\epsilon$. It is reassuring to observe that the shuffle+INV is slightly closer or comparable with the Gaussian mechanism especially when the value of $n$ is small ($n=10,000$) and the privacy level is low ($\epsilon=6$). Figure \ref{san_paris_box} (c) and (d) shows the TV distance between the estimated and the original distributions and the utility difference for locations in Paris dataset with $n$ ranging from 1,000 to 10,000 and the other parameters being the same as the experiments for the San Francisco dataset. 

\begin{table} 
\centering
\caption{Individual specific utility comparison of central and shuffle models for Gowalla data ($\epsilon=4$, $\epsilon_0=4$)}\label{table:gowalla_individual}

  \begin{tabular}{ccc|ccc}
    \hline
     {$x_0$} & 
      \multicolumn{2}{c|}{San Francisco} &     {$x_0$} & 
      \multicolumn{2}{c}{Paris} \\
    & Gaussian & shuffle+INV &  &Gaussian & shuffle+INV \\
    \hline
    40 & 4E-6 & 1E-3 & 20 & 2E-6 & 3E-4 \\
    \hline
    80 & 3E-5 & 5E-4 & 40 &  3E-5  & 2E-3 \\
    \hline
    120 & 9E-6 & 1E-3 & 60  & 4E-5 & 2E-3 \\
    \hline
    160 & 4E-5 & 2E-4 & 80 & 5E-5 & 4E-4 \\
    \hline
    200 & 2E-5 & 2E-4 & 100 & 7E-5 & 1E-4 \\
    \hline
  \end{tabular}%

\end{table}

The overall trend of TV distance for the dataset of Paris is the same as that of San Francisco. Again, we observe that the utility of the shuffle+INV is better than that of just shuffle with $k$-RR, and the utilities of the shuffle+INV and the optimal Gaussian mechanism are almost indistinguishable when the number of samples and the privacy level are low. As we see from the heatmaps in Figure \ref{san_paris_noise}, when the value of $\epsilon$ is 4, both the Gaussian mechanism and shuffle+INV generate results very close to the original distribution. Individual-specific utilities for the Paris and San Francisco datasets are described in Table \ref{table:gowalla_individual}.

\section{Conclusion}

In this paper, we have compared the privacy-utility trade-off of two different models of differential privacy for histogram queries: the classic central model with the optimal Gaussian mechanism and the shuffle model with $k$-RR mechanism as the local randomizer, enhanced with post-processing to de-noise the resulting histogram. In order to do this comparison, we needed to derive the tight bounds for the level of privacy provided by the shuffle model, so that we could tune the parameters of the Gaussian mechanism to provide the same privacy. 

First, we have used a result on the condition for tightness of ADP given by Sommer et al. in \cite{sommer2019privacy} and translated it in the context of shuffle models, giving rise to a closed form expression of the least $\delta$ for any $\epsilon$ and, thus, we obtained a necessary and sufficient condition to have the tight DP guarantee for the shuffle models. This result shows that the differential privacy ensured by the shuffle models under a certain level of local noise is much higher than what has been known by the community so far. Then, we performed experiments on synthetic and real location data from San Francisco and Paris, and we compared the statistical utilities of the shuffle and the central models. We observed that, although the central model still performs better than the shuffle model, only ever so slightly -- the gap between their statistical utilities is very small and tends to vanish as the number of samples is small.

\section{Acknowledgment} 
The work is supported by the European Research Council (ERC) project HYPATIA under the European Unions Horizon 2020 research and innovation programme. Grant agreement no. 835294 and ELSA – European Lighthouse on Secure and Safe AI funded by the European Union under grant agreement
No. 101070617.

\bibliographystyle{splncs04}
\bibliography{reference}

\appendix
\section{Proof of Theorem~Theorem~\ref{th:tightdelta_ADP}}~\label{app:proofs}

Setting $p=\mathbb{P}[x_0|x_0],\,\overline{p}=\mathbb{P}[x_0|y\neq x_0]$ in $\mathcal{R}_{\operatorname{kRR}},\,\forall s\in[n]$, $\mathbb{P}[\mathcal{M}_{x_0}(x_0)=s]$

\begin{flalign}
    {}&=p\sum\limits_{r=0}^{s-1}\left[\binom{n_{x_0}}{r}p^r(1-p)^{n_{x_0}-r}\binom{n-1-n_{x_0}}{s-1-r}\overline{p}^{s-1-r}(1-\overline{p})^{n-n_{x_0}-s+r}\right]\nonumber\\
    {}&+(1-p)\sum\limits_{r=0}^{s}\left[\binom{n_{x_0}}{r}p^r(1-p)^{n_{x_0}-r}\binom{n-1-n_{x_0}}{s-r}\overline{p}^{s-r}(1-\overline{p})^{n-n_{x_0}-1-s+r}\right]\nonumber\\
    {}&=\frac{e^{\epsilon_0}}{e^{\epsilon_0}+k-1}\sum\limits_{r=0}^{s-1}\left[\binom{n_{x_0}}{r}\frac{e^{r\epsilon_{0}}(k-1)^{n_{x_0}-r}}{(e^{\epsilon_0}+k-1)^{n_{x_0}}}\binom{n-1-n_{x_0}}{s-1-r}\frac{(e^{\epsilon_0}+k-2)^{n-n_{x_0}-s+r}}{(e^{\epsilon_0}+k-1)^{n-1-n_{x_0}}}\right]\nonumber\\
    {}&+\frac{k-1}{e^{\epsilon_0}+k-1}\sum\limits_{r=0}^{s}\left[\binom{n_{x_0}}{r}\frac{e^{r\epsilon_{0}}(k-1)^{n_{x_0}-r}}{(e^{\epsilon_0}+k-1)^{n_{x_0}}}\binom{n-1-n_{x_0}}{s-r}\frac{(e^{\epsilon_0}+k-2)^{n-n_{x_0}-1-s+r}}{(e^{\epsilon_0}+k-1)^{n-1-n_{x_0}}}\right]\nonumber\\
    {}&=\frac{e^{\epsilon_0}(k-1)^{n_{x_0}}(e^{\epsilon_0}+k-2)^{n-n_{x_0}-s}}{(e^{\epsilon_0}+k-1)^n}
    \sum\limits_{r=0}^{s-1}\binom{n_{x_0}}{r}\binom{n-1-n_{x_0}}{s-1-r}\kappa_1^r\nonumber\\
    {}&+\frac{(k-1)^{n_{x_0}+1}(e^{\epsilon_0}+k-2)^{n-n_{x_0}-1-s}}{(e^{\epsilon_0}+k-1)^n}
    \sum\limits_{r=0}^{s}\binom{n_{x_0}}{r}\binom{n-1-n_{x_0}}{s-r}\kappa_1^r\nonumber\\
    {}&=\kappa_3\left[e^{\epsilon_0}\sum\limits_{r=0}^{s-1}\binom{n_{x_0}}{r}\binom{n-1-n_{x_0}}{s-1-r}\kappa_1^r
    +\kappa_2\sum\limits_{r=0}^{s}\binom{n_{x_0}}{r}\binom{n-1-n_{x_0}}{s-r}\kappa_1^r\right]\nonumber\\
    {}&\text{Using elementary combinatorial identities, we reduce to:}\nonumber\\ 
    {}&\kappa_3\left[\kappa_2\sum\limits_{r=0}^{s}\binom{n_{x_0}}{r}\kappa_1^r\left(\binom{n-1-n_{x_0}}{s-1-r}+\binom{n-1-n_{x_0}}{s-r}\right)\right.\nonumber\\
    {}&\left.+(e^{\epsilon_{0}}-\kappa_2)\left(e^{\epsilon_0}\sum\limits_{r=0}^{s-1}\binom{n_{x_0}}{r}\binom{n-1-n_{x_0}}{s-1-r}\kappa_1^r\right)\right]\nonumber\\{}&=\kappa_3\left[\kappa_2\sum\limits_{r=0}^{s}\binom{n_{x_0}}{r}\binom{n-n_{x_0}}{s-r}\kappa_1^r
    +(e^{\epsilon_{0}}-\kappa_2)\left(\sum\limits_{r=0}^{s}\binom{n_{x_0}}{r}\binom{n-1-n_{x_0}}{s-1-r}\kappa_1^r\right)\right]\nonumber\\
    {}&=\kappa_3\left[\kappa_2\sum\limits_{r=0}^s\binom{n_{x_0}}{r}\binom{n-n_{x_0}}{s-r}\kappa_1^r
    +\frac{(e^{\epsilon_0}-\kappa_2)(s-r)}{n-n_{x_0}}\sum\limits_{r=0}^s\binom{n_{x_0}}{r}\binom{n-n_{x_0}}{s-r}\kappa_1^r\right]\nonumber\\
    {}&=\frac{\kappa_3}{n-n_{x_0}}\sum\limits_{r=0}^s\mu(s,r)\tau_r \text{ [$\mu$ and $\tau$ are as in Definition \ref{finaldelta}]}\label{eq:freqquerysame}
\end{flalign}

By similar arguments as above, for any $s\in\{0,\ldots,n\}$, $\mathbb{P}[\mathcal{M}_{x_0}(x_1)=s]$
\begin{flalign}
    {}&=\frac{1}{e^{\epsilon_0}+k-1}\sum\limits_{r=0}^{s-1}\left[\binom{n_{x_0}}{r}\frac{e^{r\epsilon_{0}}(k-1)^{n_{x_0}-r}}{(e^{\epsilon_0}+k-1)^{n_{x_0}}}
    \binom{n-1-n_{x_0}}{s-r}\frac{(e^{\epsilon_0}+k-2)^{n-n_{x_0}-s+r}}{(e^{\epsilon_0}+k-1)^{n-1-n_{x_0}}}\right]\nonumber\\
    {}&+\frac{e^{\epsilon_0}+k-2}{e^{\epsilon_0}+k-1}\sum\limits_{r=0}^{s}\left[\binom{n_{x_0}}{r}\frac{e^{r\epsilon_{0}}(k-1)^{n_{x_0}-r}}{(e^{\epsilon_0}+k-1)^{n_{x_0}}}\binom{n-1-n_{x_0}}{s-1-r}\frac{(e^{\epsilon_0}+k-2)^{n-n_{x_0}-1-s+r}}{(e^{\epsilon_0}+k-1)^{n-1-n_{x_0}}}\right]\nonumber\\
    {}&=\kappa_3\left(\sum\limits_{r=0}^{s}\binom{n_{x_0}}{r}\binom{n-1-n_{x_0}}{s-1-r}\kappa_1^r
    +\sum\limits_{r=0}^{s}\binom{n_{x_0}}{r}\binom{n-1-n_{x_0}}{s-r}\kappa_1^r\right)\nonumber\\
    {}&=\kappa_3\sum\limits_{r=0}^s\binom{n_{x_0}}{r}\binom{n-n_{x_0}}{s-r}\kappa_1^r\label{eq:freqquerydiff}
\end{flalign}
Using Result 1, for every $k>2$ and $s\in\{0,1,\ldots,n\}$, we can say that $\mathcal{M}$ induces a tight $(\epsilon,\,\delta)$-ADP guarantee with respect to $x_0,\,x_1\in\mathcal{X}$ for any $\epsilon>0$ and $\delta$ iff $\delta$ is defined as:
\begin{align}\label{freqqueryIFFCondition}
    \delta(\epsilon)=\sum\limits_{v: v>\epsilon}(1-e^{\epsilon-v})\sum\limits_{\substack{s=0\\v=\ln{\frac{\mathbb{P}[\mathcal{M}_{x_0}(x_0)=s]}{\mathbb{P}[\mathcal{M}_{x_0}(x_1)=s]}}}}^n\mathbb{P}[\mathcal{M}_{x_0}(x_0)=s]
\end{align}
Using the expressions derived for $\mathbb{P}[\mathcal{M}_{x_0}(x_0)=s]$ and $\mathbb{P}[\mathcal{M}_{x_0}(x_1)=s]$ in \eqref{eq:freqquerysame} and \eqref{eq:freqquerydiff}, respectively, to get $v_s$:

\begin{align}
&=\ln{\frac{\mathbb{P}[\mathcal{M}_{x_0}(x_0)=s]}{\mathbb{P}[\mathcal{M}_{x_0}(x_1)=s]}}=\ln{\frac{e^{\epsilon_0}\sum\limits_{r=0}^{s-1}\binom{n_{x_{0}}}{r}\binom{n-1-n_{x_0}}{s-1-r}\kappa_1^r+\kappa_2\sum\limits_{r=0}^{s}\binom{n_{x_{0}}}{r}\binom{n-1-n_{x_0}}{s-r}\kappa_1^r}{\sum\limits_{r=0}^{s-1}\binom{n_{x_{0}}}{r}\binom{n-1-n_{x_0}}{s-1-r}\kappa_1^r+\sum\limits_{r=0}^{s}\binom{n_{x_{0}}}{r}\binom{n-1-n_{x_0}}{s-r}\kappa_1^r}}\nonumber\\
&=\ln{\left(\kappa_2+\frac{(e^{\epsilon_{0}}-\kappa_2)\left(\sum\limits_{r=0}^{s-1}\binom{n_{x_0}}{r}\binom{n-1-n_{x_0}}{s-1-r}\kappa_1^r\right)}{\sum\limits_{r=0}^{s}\binom{n_{x_0}}{r}\binom{n-n_{x_0}}{s-r}\kappa_1^r}\right)}\nonumber\\
&=\ln{\left(\kappa_2+\frac{\frac{(e^{\epsilon_{0}}-\kappa_2)}{n-n_{x_0}}\left(\sum\limits_{r=0}^{s}(s-r)\binom{n_{x_0}}{r}\binom{n-1-n_{x_0}}{s-1-r}\kappa_1^r\right)}{\sum\limits_{r=0}^{s}\binom{n_{x_0}}{r}\binom{n-n_{x_0}}{s-r}\kappa_1^r}\right)}\label{eq:privacylossclosedform}
\end{align}

Combining \eqref{freqqueryIFFCondition} and \eqref{eq:privacylossclosedform}, $\delta(\epsilon)=\sum\limits_{\substack{u: u>\epsilon; s=0\\v=\ln{\frac{\mathbb{P}[\mathcal{M}_{x_0}(x_0)=s]}{\mathbb{P}[\mathcal{M}_{x_0}(x_1)=s]}}}}^n(1-e^{\epsilon-v})\mathbb{P}[\mathcal{M}_{x_0}(x_0)=s]$
\begin{align}\label{freqquery5}
    &=\sum\limits_{s=0}^n\mathbbm{1}_{\{v_{s}>\epsilon\}}(1-e^{\epsilon-v_{s}})\mathbb{P}[\mathcal{M}_{x_0}(x_0)=s]\nonumber\\
    &=\sum\limits_{s=0}^n\mathbbm{1}_{\{v_{s}>\epsilon\}}(1-e^{\epsilon-v_{s}})\frac{\kappa_3}{n-n_{x_0}}\sum\limits_{r=0}^s\mu(s,r)\tau_r\nonumber=\hat{\delta}(\epsilon)\\
    &\text{[Substituting $\mathbb{P}[\mathcal{M}_{x_0}(x_0)=s]$ from \eqref{eq:freqquerysame}].}\nonumber
\end{align}

\section{Theoretical outline}~\label{app:theory_outline}

In $\mathcal{M}$, we extend the idea of ADP to a non-adapted, general DP by using the highest value of $\delta$ across the primary inputs of every member in $\mathfrak{U}$, for a fixed $\epsilon$. This essentially ensures the worst possible tight differential privacy guarantee for the shuffle model. After that, we focus on estimating the original distribution of the primary initial dataset.

Let $\mathcal{R}_{\text{kRR}}^{-1}$ denote the inverse\footnote{the inverse of a $k$-RR mechanism always exists \cite{agrawal2005privacy, kairouz2016discrete}} of the probabilistic mechanism $\mathcal{R}_{\text{kRR}}$, which is used as the local randomizer for $\mathcal{M}$. Note that $\mathcal{R}_{\text{kRR}}^{-1}$ and $\mathcal{R}_{\text{kRR}}$ are both $k\times k$ stochastic channels as $|\mathcal{X}|=k$. Staying consistent with our previously developed notations, let us, additionally, introduce $H_{\mathfrak{N}}$ broadcasting the frequencies of the elements in $\mathcal{X}$ after they have been sanitized with $\mathfrak{N}$. In other words, $H_{\mathfrak{N}}=\mathfrak{N}_{\epsilon,\delta}(D_{\mathcal{X}})=(H_{x_0},\ldots,H_{x_{k-1}})$, where $H_{x_i}$ is the random variable giving the frequency of $x_i$ after $D_{\mathcal{X}}$ has been obfuscated with $\mathfrak{N}_{\epsilon,\delta}$.

Since both $\mathcal{M}$ and $\mathfrak{N}$ are probabilistic mechanisms, to estimate their utilities we study how accurately we can estimate the true distribution from which $D_{\mathcal{X}}$ is sampled, after observing the response of the histogram queries in both the scenarios. 

Let $\pi=(\pi_{x_0},\ldots,\pi_{x_{k-1}})$ be  the distribution of the original messages in $D(x_0)$. Our best guess of the original distribution by observing the noisy histogram going through the Gaussian mechanism is the noisy histogram itself, as $\mathbb{E}(H_{x_i})=n\pi_{x_i}$ for every $i\in\{0,\ldots,k-1\}$.

However, in the case where $D(x_0)$ is locally obfuscated using $\mathcal{R}_{\text{kRR}}$ and the frequency of each element is broadcast by the shuffle model $\mathcal{M}$, we can use the matrix inversion method \cite{agrawal2005privacy, kairouz2016discrete} to estimate the distribution of the original messages in $D(x_0)$. So $\mathcal{M}(D(x_0))\mathcal{R}_{\text{kRR}}^{-1}$ (referred as \emph{shuffle+INV} in the experiments) should be giving us $\hat{\pi}=(\hat{\pi}_{x_0},\ldots,\hat{\pi}_{x_{k-1}})$ -- the most likely estimate of the distribution of each user's message in $D(x_0)$ sampled from $\mathcal{X}$ -- where $\hat{\pi}_{x_i}$ denotes the random variable estimating the normalised frequency of $x_i$ in $D(x_0)$. 
\begin{equation}\label{shufflemodelexpect}
   \mathbb{E}(\hat{\pi})=\mathbb{E}(\mathcal{M}(D(x_0))\mathcal{R}_{\text{kRR}}^{-1})=\pi \mathcal{R}_{\text{kRR}}\mathcal{R}_{\text{kRR}}^{-1}=\pi 
\end{equation}

We recall that $\mathcal{M}$ provides tight $(\epsilon,\,\delta)$-ADP for $x_0,\,x_1$, where $\delta$ is a function of $\epsilon_0,\,\epsilon,\,\text{and }x_0$ -- essentially $\mathcal{M}$ privatizes the true query response for $x_0$ to be identified as that for any $x_1\neq x_0$. On the other hand, $\mathfrak{N}_{\epsilon,\delta}$ ensures $(\epsilon,\,\delta)$-DP, which essentially means it guarantees $(\epsilon,\,\delta)$-ADP for every $x_i\in\mathcal{X}$. 
Therefore, in order to facilitate a fair comparison of utility between the central and shuffle models of differential privacy under the same privacy level for the histogram query, we introduce the following concepts:

\begin{enumerate}
    \item[i)] Individual specific utility: Suppose the primary input of $u$ is $x_0$. \emph{Individual specific utility} refers to measuring the utility for the specific message $x_0$ in the dataset $D(x_0)$ in a certain privacy mechanism. In particular, the individual specific utility of $x_0$ in $D(x_0)$ for $\mathcal{M}$ is \begin{equation*}
        \overline{\mathcal{W}}(\mathcal{M},x_0)=|n\hat{\pi}_{x_0}-n\pi_{x_0}|,
    \end{equation*} and that for $\mathfrak{N}_{\epsilon,\delta}$ is
    \begin{equation*}
        \overline{\mathcal{W}}(\mathfrak{N}_{\epsilon,\delta},x_0)=|n\pi_{x_0}-H_{x_0}|
    \end{equation*}
    \item[ii)] Community level utility: Here we consider the utility privacy mechanisms over the entire community, i.e., all the values of the original dataset, by measuring the distance between the estimated original distribution obtained from the observed noisy histogram and the original distribution of the source messages itself. 
    
    In particular, fixing any $\epsilon_0>0$ and $\epsilon>0$, the \emph{community level utility} for $\mathcal{M}$ is 
    \begin{equation} \label{commintyutility:shuffle}
        \mathcal{W}(\mathcal{M})=d(n\hat{\pi},\,n\pi), 
    \end{equation}
    and that for $\mathfrak{\mathfrak{N}}_{\epsilon,\delta}$\footnote{where $\delta$ is correspondingly obtained using Result 1.} is
    \begin{equation}\label{communityutility:central}
        \mathcal{W}(\mathfrak{N}_{\epsilon,\delta})=d(H_{\mathfrak{N}_{\epsilon,\delta}},\,n\pi),
    \end{equation}
    where $d(.)$ is any standard metric\footnote{we consider Total Variation Distance for our experiments} to measure probability distributions over a finite space.
    
    For an equitable comparison between $\mathcal{M}$ and $\mathfrak{N}$, we take the worst tight ADP guarantee over every user's primary input and call this the \emph{community level tight DP guarantee for} $\mathcal{M}$. That is, for a fixed $\epsilon_0,\,\epsilon>0$, we have $\mathcal{M}$ satisfying $(\epsilon,\,\hat{\delta})$-DP as the community level tight DP guarantee if
    \begin{equation}\label{eq:maxdelta}
        \hat{\delta}=\max\limits_{x\in\mathcal{X}}\{\delta:\mathcal{M}\text{ is tightly }(\epsilon,\delta(x))\text{-ADP for } x\in D_{\mathcal{X}}\}
    \end{equation}
    
    Therefore, we impose the worst tight ADP guarantee on $\mathcal{M}$ over all the original messages with $\epsilon$ and $\hat{\delta}$, implying that $\mathcal{M}$ now gives a $(\epsilon,\,\hat{\delta})$-DP guarantee by Remark \ref{ADPandDP}, placing us in a position to compare the community level utilities of the shuffle and the central models of DP under the histogram query for a fixed level of privacy. In particular, we juxtapose $\mathcal{W}(\mathcal{M})$ with $\mathcal{W}(\mathfrak{N}_{\epsilon,\hat{\delta}})$, as seen in the experimental results with location data from San Francisco and Paris in Figures \ref{san_paris_box}. 
\end{enumerate}

\end{document}